\documentstyle[12pt,a4]{article}
\begin{document}
\newcommand{\bea}{\begin{eqnarray}}
\newcommand{\eea}{\end{eqnarray}}
\newcommand{\be}{\begin{equation}}
\newcommand{\ee}{\end{equation}}
\newcommand{\non}{\nonumber}
\global\parskip 6pt
\begin{titlepage}
\rightline{gr-qc/9504005}
\rightline{ITFA-95-02}
\rightline{April 1995}
\begin{center}
{\Large\bf Cobordism Effects in the Regge Calculus}\\
\vskip .25in
{\Large\bf Approach to Quantum Cosmology}\\
\vskip .50in
Danny Birmingham \footnote{Supported by Stichting voor Fundamenteel
Onderzoek der Materie (FOM)\\
Email: Dannyb@phys.uva.nl}     \\
\vskip .10in
{\em Universiteit van Amsterdam, Instituut voor Theoretische Fysica,\\
Valckenierstraat 65, 1018 XE Amsterdam, \\
The Netherlands} \\
\end{center}
\vskip .10in
\begin{abstract}
We study the ground state wave function for a universe which is
topologically a lens space within the Regge calculus approach.
By restricting the four dimensional simplicial complex to be a cone
over the boundary lens space, described by a single internal edge
length, and a single boundary edge length, one can analyze in detail
the analytic properties of the action in the space of complex edge lengths.
The classical extrema and convergent steepest descent contours of
integration yielding the wave function are found.
Both the Hartle-Hawking and Linde-Vilenkin type proposals are examined
and, in all cases, we find wave functions which predict Lorentzian
oscillatory behaviour in the late universe. A factorization property of
topology changing amplitudes within a restricted edge length class of
cone type complexes is established. The behaviour of the results under
subdivision of the boundary universe is also presented.
\end{abstract}
\end{titlepage}

\section{Introduction}

The study of simplicial approaches to the quantization of gravity
are generally divided into those based either on Regge calculus
\cite{Regge},
for reviews see \cite{Wheeler}-\cite{Hamber}, or
dynamical triangulations \cite{David}.
In the latter, one typically restricts attention to a given simplicial
topology with fixed lengths assigned to the edges, ($1$-simplices),
of the simplicial complex. The metric is then taken to be generated by
summing over various triangulations of the given topology. These
triangulations can be obtained by applying a set of local moves
to a starting complex, and with the specification of a Boltzmann weight,
one can thus simulate the quantum path integral for the topology
in question.

In the Regge calculus approach, on the other hand, the simplicial complex
which models the topology of interest is taken to be fixed, while
the squared edge length assignments become the dynamical variables.
Thus, the quantum theory is defined by a summation over edge lengths,
which serves to model the continuum integration over the metric
tensor. This approach enjoys some advantages, in particular one can
analyze with ease both classical and semiclassical issues, and such
calculations are often useful in determining the viability of any approach.
It is also possible to study these models within a minisuperspace
of edge lengths, whereby one truncates the allowed set of
dynamical variables to a smaller more manageable set, known as
simplicial minisuperspace.
Such information is not so readily extracted from the dynamical triangulation
framework.

In \cite{Hartle2}, the application of the Regge calculus approach
to quantum
cosmology was initiated through the study of the Hartle-Hawking
\cite{HarHaw}
ground state wave function. The particular example considered there was
to take the spatial universe to be topologically $S^{3}$.
The space of complex valued edge lengths was restricted to that consisting
of a single
internal edge length and a single boundary edge length.
A steepest descent contour of constant imaginary action which yielded
a convergent path integral was found. The resulting wave function
was shown to behave in the desired way for large values
of the boundary edge length, namely a Lorentzian oscillatory behaviour.
The important observation made in \cite{Hartle2} was that while
an integration over
real valued Euclidean geometries may yield a convergent result for
the path integral, it would not predict oscillatory behaviour of the wave
function in the late universe. As a result, one is obliged to study
the path integral in the space of complex valued edge lengths.
General criteria for defining the wave function of the universe
were explored in \cite{HallHar}, and explicit computations in
continuum minisuperspace models were performed in
\cite{HallLou1}-\cite{HallMy}.
A calculation in three dimensional Regge calculus was presented in
\cite{LouTuck}.

The purpose of the present investigation is to study non-trivial
topological and cobordism effects within this Regge calculus approach
to quantum cosmology.
In particular, we obtain the ground state wave function for a universe
which is topologically a lens space $L(p,1), p \geq 2$.
This is achieved by
considering the four dimensional simplicial complex to be given by
the cone over the boundary universe $L(p,1)$.
We again restrict attention to a minisuperspace consisting of a single
internal edge length and a single boundary edge length.
In fact, such spatial
universes correspond to those present in the Eguchi-Hanson
\cite{EguHan1}
and associated Gibbons-Hawking \cite{Haw,GibbHaw} series of gravitational
instantons, although their four dimensional topology is not
captured by the cone structures considered here.

We analyze the wave function of the Hartle-Hawking \cite{HarHaw}
and Linde-Vilenkin \cite{Linde}-\cite{Vilenkin1}
type, and in the cases considered we find convergent steepest descent
contours of integration which satisfy their requirements.
Once the convergent contours are identified, one can proceed to evaluate
the full path integral, or appeal to a semi-classical analysis.
Interestingly, it is noted that for larger values of $p$, classical
extrema of the Regge action are present in the Lorentzian regime for
smaller values of the bounding edge length, as compared to $S^{3}$.
In particular, we find that for $p \geq 5$, there are Lorentzian extrema for
all positive values of the boundary edge length.
Consequently, we find in such cases that no physically acceptable
Euclidean solutions are present.
In all cases, wave functions which exhibit Lorentzian oscillatory behaviour
in the late universe can be obtained.

We then turn our attention to a study of non-trivial topology changing
amplitudes within this framework. A particularly simple cobordism
whose universe is the disjoint union of a number of components can
be obtained by again taking the cone over the disjoint union of boundaries.
In this case, we restrict our analysis to a minisuperspace where each
boundary component is specified independently by a single edge length, and
with a single independent internal edge length emanating from the cone
vertex to each boundary.
A factorization property of the associated topology changing amplitude is
found, and thus the basic building blocks for such amplitudes are
the ground state wave functions discussed above.

Finally, we examine the behaviour of the results under lattice
subdivisions of the boundary universe. By appealing to the lattice
subdivision moves of Pachner \cite{Pach},
it is shown that by remaining within the initial minisuperspace,
one can readily read off the results on finer triangulations.

The outline of this article is as follows.
In the following section, we list for reference the relevant
equations in the general formalism of Regge calculus.
We then present in section $3$ the construction of simplicial complexes
for the lens spaces $L(p,1)$, due to \cite{Brehm}, and describe
some properties of their associated cones.
Section $4$ is devoted to a study of the  analytic structure of
the Regge action in the simplicial minisuperspace, and the classical
extrema are obtained. This allows us in section $5$ to investigate
the steepest descent contours in the space of complex edge lengths,
thus yielding the ground state wave function.
In section $6$, we consider the non-trivial cobordism consisting of
a boundary universe with a number of disjoint topologically distinct
lens spaces, and establish the factorization property. We then conclude
in section $7$ with a brief study of the behaviour of the results
under subdivisions of the triangulations.

\section{General Formalism}

The basic idea in the Regge calculus approach to the quantization of gravity
is to take the spacetime manifold to be modelled by a simplicial complex.
If one restricts attention to a fixed topology, then one can capture
this chosen topology by complexes which are combinatorially equivalent.
The dynamical variables in the theory are now given by the assignment
of squared edge length variables to the edges ($1$-simplices)
of the complex.
Thus, the path integration over the metric tensor is replaced
by an integration over a finite number of edge length variables.

Given a simplicial complex $M$ with boundary $\partial M$, possibly
containing several disjoint components,
the corresponding amplitude of quantum gravity will thus take the form:
\be
\Psi_{0}(s_{b}) = \int_{C} \; d\mu(s_{i})\;\exp[-I(s_{b},s_{i})]\;\;.
\label{r1}
\ee
Here, the variables $s_{b}$ specify the edge lengths of the boundary
and the integration is over the internal dynamical variables $s_{i}$.
The form of the measure $\mu$, the contour of integration $C$, along
with the action $I$, are required to complete the
specification of the model.

The Euclidean Einstein action with cosmological term for
a manifold with boundary is given by:
\be
I = - \frac{1}{16 \pi G} \int_{M} d^{4}x \sqrt{g} R +
\frac{2\Lambda}{16\pi G}\int_{M} d^{4} x \sqrt{g}
- \frac{2}{16 \pi G} \int_{\partial M} d^{3} x \sqrt{h} K \;\;,
\label{r2}
\ee
where $R$ is the scalar curvature of the metric $g$,
$\Lambda$ is the cosmological constant,
and $K$  is the extrinsic curvature scalar of the induced metric $h$ on
the boundary.
The simplicial analogue of this action is the
corresponding Regge action \cite{Regge,HarSor},
which takes the form:
\bea
I &=& -\frac{2}{l^{2}}\sum_{\sigma_{2} \subset int(M)} V_{2}(\sigma_{2})
\theta(\sigma_{2}) + \frac{2 \Lambda}{l^{2}}\sum_{\sigma_{4} \subset int(M)}
V_{4}(\sigma_{4}) \non\\
&-& \frac{2}{l^{2}}\sum_{\sigma_{2} \subset \partial M} V_{2}(\sigma_{2})
\psi(\sigma_{2})  \;\;,
\label{r3}
\eea
where the Planck length, in units where $\hbar = c = 1$,
is $l = (16 \pi G)^{1/2}$. The various terms in (\ref{r3}) are described
as follows.
The Einstein term is represented by a summation over internal $2$-simplices
$\sigma_{2}\subset int(M)$ (also known as hinges).
An internal hinge is any $2$-simplex
of the complex which contains at least one internal vertex,
and the notation
$int(M)$ is used to denote this set. The form of
the Einstein action involves the volume of the hinge $V_{2}(\sigma_{2})$
and the associated
deficit angle $\theta(\sigma_{2})$.
Similarly, the boundary term is given in terms
of the boundary $2$-simplices and their associated deficit angles
denoted by $\psi(\sigma_{2})$.
The cosmological term is simply represented as a sum over the volumes
$V_{4}(\sigma_{4})$ of the $4$-simplices $\sigma_{4}$
of the complex.  As mentioned above, the dynamical
variables in the Regge calculus approach are the edge length assignments.
Thus, the above action should be expressed in terms of these variables.
In fact, the Regge action is expressible in terms of the squared edge
lengths, and for details of the procedures involved we refer to
\cite{Regge}-\cite{Hamber}.

For ease of reference, we collect here some of the relevant formulae
which will be useful in the following.

The internal deficit angle is given by
\be
\theta(\sigma_{2}) = 2 \pi - \sum_{\sigma_{4} \supset \sigma_{2}}\theta_{d}(
\sigma_{2},\sigma_{4}) \;\;,
\label{r4}
\ee
where the summation is over all $4$-simplices containing the $2$-simplex
$\sigma_{2}$. The dihedral angle $\theta_{d}$ can then be expressed
in terms of the squared edge lengths, see below.
The corresponding deficit angle for the boundary $2$-simplex is:
\be
\psi(\sigma_{2}) = \pi - \sum_{\sigma_{4} \supset \sigma_{2}}\theta_{d}(
\sigma_{2},\sigma_{4})\;\;.
\label{r5}
\ee

It remains only to give the explicit formulae for the various volumes
and dihedral angles in terms of the squared edge lengths.
The volume of the $2$-simplex $\sigma_{2} = [i,j,k]$ is given by
\be
V_{2}([i,j,k]) = \frac{1}{2!}Det^{\frac{1}{2}}[M_{2}([i,j,k])] \;\;,
\ee
\label{r6}
where
\bea
M_{2}([i,j,k]) &=& \left( \begin{array}{cc}
s_{ij} &
\frac{1}{2}(s_{ij} + s_{ik} - s_{jk}) \\
\frac{1}{2}(s_{ij} + s_{ik} - s_{jk}) &
s_{ik}
\end{array} \right)  \;\;.
\label{r7}
\eea
Here, $s_{ij}$ is the squared edge length between the vertices $i$ and $j$.
Similarly, the volume of the $3$-simplex $[i,j,k,l]$ is:
\be
V_{3}([i,j,k,l]) = \frac{1}{3!}Det^{\frac{1}{2}}[M_{3}([i,j,k,l])]\;\;,
\label{r8}
\ee
where
\bea
M_{3}([i,j,k,l]) &=& \left( \begin{array}{ccc}
s_{ij} &
s_{ijk} &
s_{ijl} \\
s_{ijk} &
s_{ik} &
s_{ikl} \\
s_{ijl}  &
s_{ikl} &
s_{il} \end{array} \right) \;\;,
\label{r9}
\eea
and $s_{ijk} = \frac{1}{2}(s_{ij} + s_{ik} - s_{jk})$.
Also, the $4$-volume assumes the form
\be
V_{4}([i,j,k,l,m]) = \frac{1}{4!}Det^{\frac{1}{2}}
[M_{4}([i,j,k,l,m])]  \;\;,
\label{r10}
\ee
with
\bea
M_{4}([i,j,k,l,m]) =
\left( \begin{array}{cccc}
s_{ij} &
s_{ijk} &
s_{ijl} &
s_{ijm} \\
s_{ijk} &
s_{ik} &
s_{ikl} &
s_{ikm} \\
s_{ijl}  &
s_{ikl}  &
s_{il} &
s_{ilm} \\
s_{ijm}  &
s_{ikm} &
s_{ilm}  &
s_{im} \end{array} \right) \;\;.
\label{r11}
\eea
Finally, the dihedral angle of the $2$-simplex $\sigma_{2} = [i,j,k]$
is expressible as:
\bea
& &\theta_{d}([i,j,k],[i,j,k,l,m]) = \non\\
& &\cos^{-1}\{(\frac{1}{3!})^{2}Det[M_{33}([i,j,k],[i,j,k,l,m])]/
(V_{3}([i,j,k,l]) V_{3}([i,j,k,m]))\}\;\;, \non\\
\label{r12}
\eea
where
\bea
M_{33}([i,j,k],[i,j,k,l,m]) =
\left( \begin{array}{ccc}
s_{ij} &
s_{ijk}  &
s_{ijm}  \\
s_{ijk}  &
s_{ik} &
s_{ikm}\\
s_{ijl} &
s_{ikl}  &
s_{ilm}
\end{array} \right) \;\;.
\label{r13}
\eea

{}From the above relations, the explicit form of the action
for a given simplicial complex as a function
of the squared edge lengths can be obtained. In the sequel, we will be
interested in analyzing the analytic properties of this action
in the space of complex valued edge lengths, for the purposes of
computing the amplitudes of interest.

\section{Triangulations of Lens Spaces and their Cones}

Our task now is to obtain explicit triangulations for the lens spaces
which serve to model the spatial universe. Once this is achieved, one
can readily construct the corresponding cone over $L(p,1)$.

A particularly convenient construction of simplicial complexes for
lens spaces, with a small number of simplices, has been given in
\cite{Brehm}. We present here the essential ingredients in this construction.
These triangulations are interesting in their use of a small number
of simplices, and the accompanying dihedral automorphism group.
The triangulation of $L(k-2,1)$ for $k\geq 4$, is denoted by $S_{2k}$, and
it has dihedral automorphism group $D_{2k}$.  The number $N_{i}$
of $i$-simplices in each dimension is given by:
\bea
N_{0} &=& 2 k + 3 \;\;,\non\\
N_{1} &=& 2 k^{2} + 4 k + 3 \;\;,\non\\
N_{2} &=& 4 k^{2} + 4 k \;\;,\non\\
N_{3} &=& 2 k^{2} + 2 k\;\;.
\label{r14}
\eea
In particular, the resulting triangulation of $L(2,1)\cong RP^{3}$
with $11$ vertices is the smallest number possible for this
manifold \cite{Kuh}.

For $k\geq 4$, let $Z_{2k} = \{0,1,...,2k -1\}$ denote the additive set of
integers $mod\;\;2k$, and consider the permutations of that set defined by:
\bea
c(i) &=& i + 1 \;\;\;\; mod \;\; 2k \;\;,\non\\
a(i) &=& i + 2 \;\;\;\; mod \;\; 2k \;\;, \non\\
b(i) &=& 2k - 1\;\;\;\; mod \;\; 2k \;\;.
\label{r15}
\eea
We denote by $C_{2k} = gr(c)$ the group generated by the permutation $c$,
and its action can be naturally extended to the set of all simplices
with vertices taking values in the set $Z_{2k}$.
The automorphism group of the triangulation is the dihedral group generated
by $a$ and $b$, although for the purpose of obtaining
the simplicial complex, only the elements $c$ and $a$ are required.

The first step in the construction is to introduce the set of
$3$-simplices
\be
\Delta_{i} = [0,1,k-i,k-i+1]  \;\;\;\; with\;\; i = 0,1,...,k-3\;\;,
\label{r16}
\ee
and the corresponding orbit under the action
of $C_{2k}$ is written as $C_{2k}(\Delta_{i})$. Each
of these orbits contains $2k$ elements, save for
$C_{2k}(\Delta_{0})$ which contains $k$ elements.
A simplicial complex containing $k(2k -5)$ $3$-simplices is now given by:
\be
H_{2k} = C_{2k}(\Delta_{0}) \cup \cdots \cup C_{2k}(\Delta_{k-3}) \;\;.
\label{r17}
\ee
Here, we use the $\cup$ notation to describe the union of simplices,
but when the need arises we will specify precisely the relative
orientations of the $3$-simplices involved.

Consider now the following collection of simplices:
\bea
F_{1} &=& [0,1,3] \;\;,\non\\
F_{2} &=& [0,2,3] \;\;,\non\\
F_{3} &=& [1,2,4] \;\;,\non\\
F_{4} &=& [1,3,4] \;\;,\non\\
E_{1} &=& [0,2]   \;\;,\non\\
E_{2} &=& [1,3]   \;\;.
\label{r18}
\eea
The next step is to introduce three additional vertices
$(x,y,z)$, and consider the simplicial complex
\bea
K_{2k} &=& [y,z, C_{k}(E_{1})]\cup [x,z,C_{k}(E_{2})] \cup
[x,y,C_{k}(E_{2})] \cup \non\\
&\cup& [y,C_{k}(F_{1})] \cup [y,C_{k}(F_{2})]\cup
[z,C_{k}(F_{3})] \cup [z,C_{k}(F_{4})] \;\;,
\label{r19}
\eea
where the subgroup $C_{k} \subset C_{2k}$ of index two is given by
$C_{k} = gr(a)$.

Finally, we obtain the simplicial complex
\be
S_{2k} = H_{2k} \cup K_{2k}\;\;.
\label{r20}
\ee
According to \cite{Brehm}, $S_{2k}$ is a $D_{2k}$-symmetric triangulation
of the lens space $L(k-2,1)$, for $k\geq 4$.

One can proceed and obtain the explicit set of $3$-simplices for
a given case of interest, and we list here the resulting triangulation
of $L(2,1)$. It is important to note that the relative orientations
of the $3$-simplices in $S_{2k}$ need to be
specified in order to ensure a vanishing boundary. This yields:
\bea
L(2,1) &=& \non\\
&+& [0,1,4,5]+ [1,2,5,6]+ \;[2,3,6,7]- [0,3,4,7]+ [0,1,3,4]+ \non\\
&+& [1,2,4,5]+ [2,3,5,6]+ \;[3,4,6,7]- [0,4,5,7]+ [0,1,5,6]+\non\\
&+& [1,2,6,7]- [0,2,3,7]+ \;[y,z,0,2]+ [y,z,2,4]+ [y,z,4,6]+\non\\
&-& [y,z,0,6]- [x,z,1,3]- [x,z,3,5]- [x,z,5,7]+ [x,z,1,7]+\non\\
&+& [x,y,1,3]+ [x,y,3,5]+ [x,y,5,7]- [x,y,1,7]+ [y,0,1,3]+\non\\
&+& [y,2,3,5]+ [y,4,5,7]\;+ [y,1,6,7]- [y,0,2,3]- [y,2,4,5]+\non\\
&-& [y,4,6,7]- [y,0,1,6]\;+ [z,1,2,4]+ [z,3,4,6]+ [z,0,5,6]+\non\\
&+& [z,0,2,7]- [z,1,3,4]\;- [z,3,5,6]- [z,0,5,7]- [z,1,2,7] \;\;.\non\\
\label{r21}
\eea

Given these simplicial complexes for the spatial universe, we can
turn our hand to constructing a four-dimensional simplicial complex
which has $L(p,1)$ as its boundary.  One means of achieving this is
to consider the complex known as the cone over $L(p,1)$ \cite{Munkres}.
This simply
involves the addition of a single extra vertex, the cone vertex denoted
by $c$, and joining this to all vertices of the bounding lens space.
Each $4$-simplex of the cone then takes the form  $[c,\sigma_{3}]$,
where $\sigma_{3}$ is a $3$-simplex of the boundary complex.
It is typical to denote by $M_{4} = c\star M_{3}$ the
four dimensional cone complex
over its boundary $M_{3}$. With this orientation, the boundary
of the cone $M_{4}$ is $+M_{3}$.
The number of $i$-simplices contained in the cone are immediately
evident:
\bea
N_{0}(M_{4}) &=& N_{0}(M_{3}) + 1 \;\;,\non\\
N_{1}(M_{4}) &=& N_{1}(M_{3}) + N_{0}(M_{3}) \;\;, \non\\
N_{2}(M_{4}) &=& N_{2}(M_{3}) + N_{1}(M_{3}) \;\;, \non\\
N_{3}(M_{4}) &=& N_{3}(M_{3}) + N_{2}(M_{3}) \;\;,\non\\
N_{4}(M_{4}) &=& N_{3}(M_{3}) \;\;.
\label{r22}
\eea

As verified in \cite{Brehm}, the above triangulations of lens spaces
are in fact simplicial manifolds, satisfying the so-called
manifold condition. Given an $n$ dimensional simplicial complex $K$,
we recall that the
star of a simplex $\sigma$ in $K$ is the collection
of simplices which contain $\sigma$, together with all their
subsimplices. The link of the simplex $\sigma$ is then the set of
simplices in the star of $\sigma$ which do not contain $\sigma$.
The simplicial complex $K$ is said to be a simplicial $n$-manifold
if and only if the link of every $k$-simplex is combinatorially
equivalent to an $(n-k-1)$-sphere \cite{RS}.

In particular then, because of the manifold condition, the Euler
character of the lens space triangulations vanishes,
$N_{0} - N_{1} + N_{2} - N_{3} = 0$. The fact that
the complex is closed implies that $N_{2} = 2 N_{3}$; the $3$-simplices
are glued together pairwise. One can then
express the information in (\ref{r22}) in terms
of the two independent quantities $N_{1}(M_{3})$
and $N_{3}(M_{3})$, for example.

\section{Analysis of the Regge Action and its Extrema}

Having chosen our four dimensional spacetime to be represented by
a cone over the bounding lens space $M_{3} = L(p,1)$,
we can now obtain the explicit
form of the associated Regge action.
The cone structure introduced is appealing in the sense that we can
immediately identify a natural minisuperspace in which to
study the model. Since the only internal vertex is the cone vertex,
all the internal $1$-simplices are of the form $[c,b]$, where
$b$ is a vertex lying in the boundary. The truncation of edge length
variables to the minisuperspace of interest can be effected by
considering all internal edges
to be described by a single internal edge length, denoted $s_{i}$.
In addition, we assume that the boundary lens space is described in terms
of a single bounding edge length $s_{b}$.

As a result,
the Regge action will be a function of only two variables,
and our task will be to perform the integration over the
single internal edge length. In this respect, we will be able
to appeal to the analysis performed for the case of $S^{3}$
boundary in \cite{Hartle2}.
It is perhaps worth noting that in the
model discussed in \cite{Hartle2}, the simplicial complex with
$S^{3}$ boundary was itself a simplicial manifold. In the case under
study here, the  four-dimensional cone does not satisfy the manifold
condition, in particular because of the fact
that the link of the cone vertex is a lens space rather than $S^{3}$.
Nevertheless, it does provide a particularly simple cobordism
with a boundary of non-trivial topology. Furthermore,
the general framework of Regge calculus requires only that
spacetime be represented by a simplicial complex.

It is convenient to introduce the scaled variables:
\be
z = \frac{s_{i}}{s_{b}} \;\;, \;\;\; S = \frac{H^{2} s_{b}}{l^{2}}\;\;,
\label{r23}
\ee
where $H^{2} = l^{2}\Lambda/3$.
The general formulae of section $2$ now yield the following results.
For the cosmological term, we note that each
$4$-simplex is of the form $[c,\sigma_{3}]$, where $\sigma_{3}$
is a $3$-simplex in the boundary, and has $4$-volume:
\be
V_{4} = \frac{1}{24\sqrt{2}}s_{b}^{2} (z - \frac{3}{8})^{1/2}\;\;.
\label{r24}
\ee
The number of $4$-simplices is of course $N_{3}(M_{3})$.

To evaluate the Einstein term, we observe that
the internal $2$-simplices are all of the type $[c,\sigma_{1}]$,
where $\sigma_{1}$ is a $1$-simplex on the boundary. Hence,
there are $N_{1}(M_{3})$ internal $2$-simplices each with volume:
\be
V_{2}(int)= \frac{s_{b}}{2} (z - \frac{1}{4})^{1/2}\;\;.
\label{r25}
\ee
The associated dihedral angle of this internal $2$-simplex is then
\be
\theta_{d}(int) =   \cos^{-1}[\frac{(2z - 1)}{2(3z - 1)}] \;\;.
\label{r26}
\ee
Here, one uses the fact that the volumes of the internal and boundary
$3$-simplices are:
\bea
V_{3}(int) &=& \frac{1}{12}s_{b}^{3/2}(3z - 1)^{1/2}\;\;,\non\\
V_{3}(bound) &=& \frac{\sqrt{2}}{12} s_{b}^{3/2}\;\;.
\label{r26b}
\eea

Turning now to the boundary term in the action, we again have
a single type of $2$-simplex. There are $2N_{3}(M_{3})$ of these
and the volume of each is:
\be
V_{2}(bound) = \frac{\sqrt{3}}{4} s_{b} \;\;.
\label{r27}
\ee
The dihedral angle in this case is given by
\be
\theta_{d}(bound) = \cos^{-1}[\frac{1}{2\sqrt{2}(3z - 1)^{1/2}}]\;\;.
\label{r28}
\ee

Assembling the various terms, we obtain the complete Regge
action in the form
\be
I(z,S) = [-S F(z) + S^{2} G(z)]/H^{2}  \;\;,
\label{r29}
\ee
where
\be
F(z) = a_{1}[\pi - 2\cos^{-1}(z_{1})] + a_{2}(z - \frac{1}{4})^{1/2}
[2 \pi - a_{3}\cos^{-1}(z_{2})]     \;\;,
\label{r30}
\ee
\be
G(z) = a_{4}(z - \frac{3}{8})^{1/2} \;\;,
\label{r31}
\ee
and we have introduced the variables
\be
z_{1} = \frac{1}{2\sqrt{2}(3z - 1)^{1/2}} \;\;\;, \;\;
z_{2} = \frac{(2z - 1)}{2(3z -1)}\;\;.
\label{r32}
\ee
The coefficients $a_{i}$ appearing in the action are expressed in terms
of the number of $i$-simplices of the boundary as follows:
\bea
a_{1} &=& N_{3}(M_{3}) \sqrt{3} \;\;,\non\\
a_{2} &=& N_{1}(M_{3}) \;\;,\non\\
a_{3} &=& 6 N_{3}(M_{3})/N_{1}(M_{3}) \;\;,\non\\
a_{4} &=& N_{3}(M_{3})/(4\sqrt{2})  \;\;.
\label{r33}
\eea
The values are collected in Table $1$ for the cases $ 2 \leq p \leq 7$,
and can be read off from equation (\ref{r14}) in general.

\begin{table}
\begin{center}
\begin{tabular}{|c|c|c|c|c|c|} \hline \hline
 &  &  &  &  & \\
$M_{3}$ & $a_{1}$ & $a_{2}$ & $a_{3}$ & $a_{4}$ & $S_{crit}$ \\ \hline
 &  &  &  &  & \\
$L(2,1)$ & $40\sqrt{3}$ & $51$ & $240/51$ & $5\sqrt{2}$ &
$3.53726$ \\ \hline
 &  &  &  &  & \\
$L(3,1)$ & $60\sqrt{3}$ & $73$ & $360/73$ & $15/\sqrt{2}$ &
$1.46391$ \\ \hline
 &  &  &  &  & \\
$L(4,1)$ & $84\sqrt{3}$ & $99$ & $504/99$ & $21/\sqrt{2}$ &
$0.109891$ \\ \hline
 &  &  &  &  & \\
$L(5,1)$ & $112\sqrt{3}$ & $129$ & $672/129$ & $14\sqrt{2}$ &
$-0.842155$ \\ \hline
 &  &  &  &  & \\
$L(6,1)$ & $144\sqrt{3}$ & $163$ & $864/163$ & $18\sqrt{2}$ &
$-1.54737$ \\ \hline
 &  &  &  &  & \\
$L(7,1)$ & $180\sqrt{3}$ & $201$ & $1080/201$ & $45/\sqrt{2}$
& $-2.09039$ \\ \hline \hline
\end{tabular}
\end{center}
\vspace{0.5 cm}
\caption{The coefficients $a_{i}$ appearing in the Regge action for
the cone over the lens space $L(p,1)$,
and the critical value $S_{crit}$ of the boundary edge length.}
\end{table}

It is important to make the following
observation regarding the factor of $2$ appearing in the
formula for deficit angle of a boundary $2$-simplex (the first term
in $F(z)$).
The bounding lens space is represented by a closed simplicial
complex, and closure of the complex means that each $2$-simplex is
contained in  precisely two $3$-simplices.
Thus, when we elevate this boundary
complex to its associated cone, we immediately know that the
number of
$4$-simplices containing each boundary $2$-simplex is again precisely
$2$, each being of the form $[c,\sigma_{3}]$, with $\sigma_{3}$
belonging to the boundary.
This fact becomes crucially relevant when we search for the extrema of
the Regge action.

The value of the coefficient $a_{3}$ is obtained by
determining the number of $4$-simplices which contain each internal
$2$-simplex. Since each internal $2$-simplex is of the form $[c,\sigma_{1}]$,
where $\sigma_{1}$ is a $1$-simplex in the boundary, we must equivalently
determine the number of boundary $3$-simplices containing
a given boundary $1$-simplex. This number depends on the individual
$1$-simplex. However, since each $3$-simplex contains six $1$-simplices,
the sum of the number of $3$-simplices containing
all the $1$-simplices is clearly $6 N_{3}(M_{3})$. Therefore, the
sum of the internal deficit angles takes the form given above,
with the quoted values of  $a_{2}$ and $a_{3}$.

The explicit form of the Regge action enables us to
examine its analytic and asymptotic properties. On general grounds,
as noted in \cite{Hartle2}, the action is an analytic function of the
squared edge lengths, apart from the presence of certain branch surfaces
and logarithmic infinites, which arise when
the volumes of various simplices vanish.
In the case at hand, we can explicitly
identify the analytic nature of the action $I(z,S)$. In fact,
because of our choice of cone complexes and simplicial minisuperspace, the
analytic behaviour
of the action parallels that presented in \cite{Hartle2} for
the case of $S^{3}$.

There is a square root branch point at $z = 3/8$, where the volume of
the $4$-simplices vanishes. Similarly, one spots a square root branch
point when the volume of the  internal $2$-simplices vanishes, at $z = 1/4$.
In addition, the vanishing of the volume of internal $3$-simplices
is  responsible for  a square root branch point at $z = 1/3$; this point
also affords a logarithmic branch point. Using the representation,
\be
\cos^{-1}(z) = -i\;\log[z + \sqrt{z^{2} -1}]\;\;,
\label{r34}
\ee
we find that in the neighbourhood of
$z=1/3$, the action has the behaviour:
\be
I(z,S) \sim \frac{i}{H^{2}}\;S\;(2\sqrt{3} N_{3}(M_{3}))\;\log[3z -1] \;\;.
\label{r35}
\ee

According to equation (\ref{r11}) the metric
in each $4$-simplex is specified in terms of the
squared edge lengths, and is real for real
values of $s_{ij}$. In our specific minisuperspace, we have just
a single type of $4$-simplex, and
the signature of the metric depends on the value
of the variable $z$, the corresponding eigenvalues being
\be
\lambda = (\frac{1}{2},\frac{1}{2},\frac{1}{2},4(z - \frac{3}{8}))\;\;.
\label{r36}
\ee
Thus, for real $z > \frac{3}{8}$, we have a regime of real Euclidean
geometries, while for real $z < \frac{3}{8}$ there lies a regime
of real geometries of Lorentzian signature.

Due to the presence of the branch points, we need to declare the location
of the branch cuts, and the corresponding phases of the action
on its various sheets will
then be determined. One first notes that the function $\cos^{-1}(z)$
has branch points at $-1$, $+1$, and $\infty$, and conventionally
the branch cuts are placed from $-\infty$ to $-1$, and from
$+1$ to $+ \infty$. With this choice, $\cos^{-1}(z)$ is real
for real $-1 < z < +1$.
The corresponding cuts for $\cos^{-1}(z_{2})$ in the $z$-plane
then lie between the
points $\frac{1}{3}$ to $\frac{3}{8}$, and $\frac{1}{4}$ to $\frac{1}{3}$,
respectively.
The branch cuts of $\cos^{-1}(z_{1})$ lie between $\frac{1}{3}$ and
$\frac{3}{8}$. In addition, because of the presence of the square
root branch points at $\frac{1}{4}$, $\frac{1}{3}$, and $\frac{3}{8}$,
it is convenient to define a first sheet for the action $I(z,S)$
which a branch cut extending from $\frac{3}{8}$ to $- \infty$.

With the $z$-plane cut in the above way, we note that for real
$z > \frac{3}{8}$, we have real valued Euclidean signature action,
with real volumes and real deficit angles. The action on the first sheet
is given by eqns. (\ref{r29})-(\ref{r32}) with positive signs taken for the
square root factors.

As we have seen, for real $z < \frac{3}{8}$, we have a region
of Lorentzian geometries. In particular, however, in the range
real $z < \frac{1}{4}$, we find that the action is purely imaginary.
On the first sheet, we have
\bea
F(z) &=& i\; a_{1}[- 2\sinh^{-1}(\frac{1}{2 \sqrt{2}(1 - 3z)^{1/2}})]
\non\\
&+& i\; a_{2}\;(\frac{1}{4} - z)^{1/2}
[2 \pi - a_{3}\cos^{-1}(\frac{(2z -1)}{2(3z - 1)})]     \;\;,
\label{r37}
\eea
and
\be
G(z) = i \;a_{4}\;(\frac{3}{8} - z)^{1/2} \;\;.
\label{r38}
\ee
The identity
\be
\pi - 2 \cos^{-1}(i z) = 2 \sin^{-1}(iz) = 2 i\;\sinh^{-1}(z)\;\;,
\label{r38b}
\ee
has been used in the above. It is here that we notice that the factor
of $2$ in the boundary deficit angle is crucial, so that the action in the
range real $z < \frac{1}{4}$
is purely imaginary. As we shall see, this is important
for the existence of Lorentzian signature solutions to the Regge equations
of motion.
One now sees that if the function is continued once around all the branch
points at $z=\frac{1}{4}$, $\frac{1}{3}$, and $\frac{3}{8}$, we will reach
a second sheet, and the value of the action is the negative of its value
on the first sheet. Thus, continuing twice around all branch points
returns the function to its initial value.  This behaviour can be
established by simply noting
that encircling all the branch points renders no change in
$\cos^{-1}(z_{2})$, while $\cos^{-1}(z_{1}) \rightarrow \cos^{-1}(-z_{1}) =
\pi - \cos^{-1}(z_{1})$.

The asymptotic behaviour of the action is important when
discussing the extrema of the action, and when searching for
convergent contours of integration.
For large $|z|$ on the first sheet, we have
\be
F(z) \sim a_{4}\; S_{crit}\; (z- \frac{1}{4})^{1/2}\;\;,
\label{r39}
\ee
and
\be
G(z) \sim a_{4}\;(z - \frac{3}{8})^{1/2}  \;\;,
\label{r40}
\ee
where
\be
S_{crit} = \frac{a_{2}}{a_{4}}[2\pi - a_{3}\cos^{-1}(\frac{1}{3})]\;\;.
\label{r41}
\ee
The asymptotic behaviour of the complete action on the first sheet
then takes the form:
\be
I(z,S) \sim \frac{a_{4}}{H^{2}}\;S(S - S_{crit})\; z^{1/2}\;\;.
\label{r42}
\ee
Thus, it is clear that the behaviour of the action is crucially dependent
on whether the value of the boundary edge length is greater or less
than the critical value, denoted by $S_{crit}$. The values of $S_{crit}$
for the cases $2 \leq p \leq 7$ are listed in Table $1$.
Using equation (\ref{r14}), one sees that $S_{crit}$ remains negative
for all $p \geq 5$, and tends to a value $S_{crit} = - 6.23708$ as $p$
approaches infinity.

We can turn now to a description and analysis of the classical
extrema of the action. The Regge equation of motion in this
model takes the simple form:
\be
\frac{d}{dz} I(z,S) = 0\;\;.
\label{r43}
\ee
This equation is to be solved for the value of $z$ subject to fixed
boundary data $S$, and via equation (\ref{r23}),
the solution then determines
a complete simplicial geometry. Our physical restriction on the chosen
boundary data is that $S$ should be real valued and positive.
The equation of motion can be rewritten in the form:
\be
S = \frac{F^{\prime}(z)}{G^{\prime}(z)} =
\frac{a_{2}}{a_{4}}\frac{(z - \frac{3}{8})^{1/2}}{(z -
\frac{1}{4})^{1/2}}[2\pi - a_{3}\cos^{-1}(z_{2})]  \;\;,
\label{r44}
\ee
where the prime indicates a derivative with respect to $z$.
It should be pointed out that the above form requires the relation
$a_{1}\sqrt{3} = a_{2}a_{3}/2$, which is the case for the models considered
here, as can be seen from equation (\ref{r33}).

Since $S$ is real and positive, solutions can exist when $F^{\prime}$
and $G^{\prime}$ are both purely real valued, or when both are purely
imaginary. Appealing to the analysis above, we can then declare
that classical extrema exist in two regions, for real $z > \frac{3}{8}$,
and for real $z < \frac{1}{4}$. The physical acceptability of such
solutions will require in addition that $S$ is positive.

{}From (\ref{r44}), we see that the value
of the boundary edge length is always $S < S_{crit}$, for
real $z > \frac{3}{8}$. Thus, for every  $0 < S < S_{crit}$,  there
exists a real Euclidean solution at real $z > \frac{3}{8}$.
Similarly, for every positive $S$ with $S > S_{crit}$,
there is a real Lorentzian
solution at real $z < \frac{1}{4}$.
It should also be noted that these solutions occur in pairs, in addition
to those on the first sheet, there are corresponding solutions on
the second sheet with opposite value of the action.
Interestingly, as can be seen from the values of $S_{crit}$ quoted
in Table $1$, real solutions with Lorentzian signature exist
for smaller values of the boundary edge length, as compared to
the case of $S^{3}$ studied in \cite{Hartle2}.

Furthermore, the value of $S_{crit}$ is negative
for $p \geq 5$. Therefore, it appears that no
Euclidean solutions exist in these models.
However, one could contemplate encircling the branch points either at
$z = \frac{3}{8}$, or  $\frac{1}{4}$.
This will cause a change of sign in one of the
square root prefactors in (\ref{r44}). However, in addition
the value of $\cos^{-1}(z_{2})$ will flip sign, and thus $S$ will
remain negative and less that $S_{crit}$. Thus, no physically
acceptable Euclidean solutions exist in these cases.

On a related matter, one should perhaps note that even in the case
when $S_{crit}$ is positive, Euclidean solutions do not exist for
every value of $z > \frac{3}{8}$. This can be seen even in the
case with  $S^{3}$  boundary. At values of $z$ close to $\frac{3}{8}$,
for example $\frac{3}{8} < z < 0.4$ for $S^{3}$,
and $\frac{3}{8} < z < 0.889512$
for $L(2,1)$,
one finds that $a_{3} \cos^{-1}(z_{2}) > 2 \pi$, resulting in
a negative value of $S$.

The structure of the classical solutions are presented for the cases
of $L(2,1)$ and $L(5,1)$ in Figures $1$ and $2$.

\section{Steepest Descent Contours for the Wave Function of the Universe}

Armed with the classical solutions to the Regge equation of motion, we
can proceed with our determination of the ground state wave
function of the universe.
The remaining data needed is a specification of the measure $\mu$, and the
integration contour $C$. We shall take the measure in the form
\be
d\mu(s_{i}) = \frac{d s_{i}}{2 \pi i l^{2}}\;\;.
\label{r45}
\ee
Therefore, the ground state wave function is given as:
\be
\Psi_{0}(S) = \frac{S}{2 \pi i H^{2}} \int_{C} dz \; \exp[-I(z,S)]\;\;.
\label{r46}
\ee

We wish to determine if $\Psi_{0}(S)$ can be obtained in a form which
exhibits oscillatory behaviour for large values of the bounding edge
length $S$. As shown in \cite{Hartle2} for a universe with $S^{3}$
topology, this is indeed possible. We recall
that there are two Lorentzian extrema when $S>S_{crit}$, one on the first
sheet of the action, and one on the second sheet.
At the corresponding values of real $z < \frac{1}{4}$, the action
is purely imaginary and of opposite sign on the two sheets.
The aim is then to identify a convergent steepest
descent contour $C$ of constant
imaginary action which passes through one, or both, of these
Lorentzian extrema.

As we have noted, the critical value $S_{crit}$ for the lens spaces
studied here
is smaller that the corresponding value for spherical topology.
Therefore, once we have succeeded in identifying the steepest descent
contours, it means we can define an oscillating wave
function for relatively small values of the bounding edge length.
Indeed, we see that for $p \geq 5$, oscillatory behaviour can be
obtained for all positive values of $S$.

Based on our explicit knowledge of the analytic properties of the action
and its extrema, we can determine the nature of the steepest descent
contour by resorting to general argument, as reviewed in \cite{Hartle2}.
Consider for example the Lorentzian extremum lying at some value of real
$z < \frac{1}{4}$; the  action here is purely imaginary  with value
$Im(I) = I_{ext}$.  By construction, a contour
of constant imaginary action consists of two
sections, one of steepest ascent, the other of
steepest descent.
Descending most steeply away from the extremum,
one could in general end either at infinity,
a singular point of the action, or at another
extremum with the same value of $Im(I)$. From (\ref{r35}), we see that the
only singular point is at $z = \frac{1}{3}$,
where the value of $Im(I)$ diverges, so the contour cannot end there.
Since the other extremum at this value of $z$ lies on the second
sheet, it has opposite value of $Im(I)$, so a contour cannot connect them.
Thus, on general grounds, the steepest descent contour
must be in infinite in extent, and
passes from infinity to infinity through the extremum.
Indeed, one can verify this explicitly,
and an example for the universe with $RP^{3}$ topology is
given in Figure $3$.

The convergence of this contour can be verified by recalling the asymptotic
behaviour of the action, as presented in the previous section.
Beginning on the first sheet, at the Lorentzian extremum with real
$z < \frac{1}{4}$, the locus of the contour on the upper
half of this sheet is given asymptotically by
\be
\frac{a_{4}}{H^{2}} S(S - S_{crit}) Im(z^{1/2}) = I_{ext} \;\;.
\label{r47}
\ee
The asymptotic behaviour of the real part of the action on this section
of the contour is
\be
Re[I(z,S)] \sim \frac{a_{4}}{H^{2}} S(S- S_{crit}) |z|^{1/2}\;\;,
\label{r48}
\ee
thus guaranteeing convergence.

Moving downwards from the extremum, we immediately cross the branch cut,
and hence pass onto the second sheet. However, due to the alteration
in sign of the action, one cannot proceed to infinity on the second sheet.
Instead, one finds that the contour enjoys traversing the branch cut
once more, moving onto a third sheet. For large values of $S$, this
crossing point lies between $z=\frac{1}{4}$ and $z = \frac{1}{3}$.
Having emerged onto the third sheet, the contour is asymptotic to
\be
\frac{a_{4}}{H^{2}} S(S + \tilde{S}_{crit}) Im(z^{1/2}) = I_{ext}\;\;,
\label{r49}
\ee
where
\be
\tilde{S}_{crit} = \frac{a_{2}}{a_{4}}[2\pi +
a_{3}\cos^{-1}(\frac{1}{3})]\;\;.
\label{r50}
\ee
Convergence is ensured due to the fact that
\be
Re[I(z,S)] \sim \frac{a_{4}}{H^{2}} S( S + \tilde{S}_{crit})|z|^{1/2}\;\;.
\label{r51}
\ee
For smaller values of $S$, the crossing point to the third sheet lies between
$z = \frac{1}{3}$ and $z=\frac{3}{8}$. When the dust settles, however,
the end result is that a convergent contour of integration for the wave
function exists. Since it passes through the Lorentzian extrema,
the desired oscillatory behaviour in the late universe, i.e.
when $S > S_{crit}$, is guaranteed. In fact the steepest descent contour
just described has a complex conjugate partner with $Im(I) = - I_{ext}$,
where one begins at the extremum on the second sheet.

It turns out that one can also find a steepest descent contour
when $S<S_{crit}$. Beginning at a Euclidean extremum with real
$z > \frac{3}{8}$ on the first sheet, the action is purely real,
and hence a contour of constant $Im(I) = 0$ is required.
Clearly, one contour extends along the real axis from $\frac{3}{8}$
to $+ \infty$, corresponding to an integration over real Euclidean
geometries. However, such a contour is of steepest ascent.
The section of steepest descent can be found, and in fact is one which
encircles all three finite branch points.
The contours for the extremum on the second sheet are equally given.

The existence of convergent steepest descent contours allows us to
perform a complete numerical integration yielding
the wave function. For illustrative purposes however, it suffices
to resort to the semi-classical approximation, and evaluate the wave
function to first order.

To implement the Hartle-Hawking proposal, we wish to obtain a real valued
wave function. This can be achieved by combining the two
sections of the contour, passing through
both Lorentzian, and both Euclidean extrema.
The form of the  wave function for $S>S_{crit}$ is therefore
\be
\Psi_{0}(S) = [\frac{S^{2}}{2 \pi H^{4} \tilde{I}^{\prime\prime}_
{ext}(S)}]^{1/2}
\;2\cos[\tilde{I}_{ext}(S) - \frac{\pi}{4}]\;\;,
\label{r52}
\ee
where the action is written as $I = i\;\tilde{I}$, and
$\tilde{I}_{ext}(S)$ and
$\tilde{I}^{\prime\prime}_{ext}(S)$ are evaluated at
the value of $z$ corresponding to the extremum $S$. The fluctuation
term, in a form valid for real $z < \frac{1}{4}$, is given by
\bea
\tilde{I}^{\prime\prime}(z,S) &=&  -\frac{i}{H^{2}}[-SF^{\prime\prime} +
S^{2}G^{\prime\prime}]\non\\
&=&\frac{S}{H^{2}}\{ \frac{a_{2}}{4}\frac{1}{(\frac{1}{4} - z)^{3/2}}[2 \pi
- a_{3}\cos^{-1}(z_{2})] + \non\\
&+& \frac{a_{2}a_{3}}{8\sqrt{2}} \frac{1}{(\frac{1}{4} - z)(\frac{3}{8}
-z)^{1/2}(1- 3z)}\} \non\\
&-& \frac{S^{2}}{H^{2}} \frac{a_{4}}{4}\frac{1}{(\frac{3}{8}
- z)^{3/2}}    \;\;.
\label{53}
\eea

For $S< S_{crit}$, the Hartle-Hawking wave function is also given
by considering the two Euclidean extrema. However, because of
the alternate signs, it is the extremum on the first sheet which
gives the dominant contribution, namely
\be
\Psi_{0}(S) = -[- \frac{S^{2}}{2 \pi H^{4} I^{\prime\prime}_{ext}(S)}]^{1/2}
\exp[- I_{ext}(S)]\;\;.
\label{r54}
\ee

It is now a simple matter to plot these wave functions for the
lens space universe. In Figure $4$, the Hartle-Hawking wave function
is given for $RP^{3}$, in the semi-classical approximation.
Correspondingly, in Figure $5$, the case of $L(5,1)$ is presented.
Here, we again remark that  an oscillating wave function is allowed
for all positive values of the boundary edge length, as is also the
case for all $p > 5$.
For all $L(p,1), p \geq 2$, we find an oscillating wave function
for large values of the boundary edge length.

Finally, we mention the wave functions of the Linde-Vilenkin
\cite{Linde,Vilenkin1}, variety.
The proposal here is to define the wave function to consist purely
of outgoing waves. In other words, the wave function should be defined
in terms of the contour passing through the Lorentzian extremum
on the first sheet alone, giving it a form:
\be
\Psi_{0}(S) = [\frac{S^{2}}{2 \pi H^{4} \tilde{I}^{\prime\prime}_
{ext}(S)}]^{1/2}
\;\exp[-i(\tilde{I}_{ext}(S) - \frac{\pi}{4})]\;\;.
\label{r54b}
\ee

Clearly then, wave functions satisfying the requirements of
the Linde-Vilenkin proposal are possible for all the lens spaces
studied here.

\section{Factorization of Topology Changing Amplitudes}

The cobordism structures that we have been discussing so far have
involved a boundary consisting of a single component $M_{3} = L(p,1)$.
The corresponding cone $M_{4} = c\star M_{3}$ over this boundary
then yields the wave function
of a universe with simplicial topology specified by $M_{3}$.
However, we may also be interested in studying
the situation when the universe
is given by the disjoint union of a number of components of varying
topology. The wave function of the universe in this case is then
a topology changing amplitude.
In general however, such an amplitude would be quite difficult to calculate.

Our aim now is to study the nature of these topology changing amplitudes
within this Regge calculus approach.
Based on our previous experience, we shall construct a simple cobordism
which reduces these amplitudes to a tractable form.

Let us suppose that the universe
consists of the disjoint union of a number of components, denoted by
$M_{3} = \cup_{j}M_{3}^{(j)}, \; (j = 1, ..., N)$.
We then construct the cone over this boundary, $M_{4} = c\star M_{3}$,
which of course has boundary $\partial M_{4} = M_{3}$.

However, given our cone structure, we can immediately identify a convenient
simplicial minisuperspace in which to study these amplitudes.
Again, we let the simplicial geometry of each boundary component
be described in terms of a single edge length $S^{(j)},\; (j=1,...,N)$.
In addition, we allow an independent internal edge length
$z^{(j)},\; (j = 1,...,N)$ to emanate
from the cone vertex to each of the boundary components.
As before, we use the scaled variables,
$z^{(j)} = s_{i}^{(j)}/s_{b}^{(j)}$ and $S^{(j)} = H^{2}s_{b}^{(j)}/l^{2}$.

The crucial observation now is that with this choice of minisuperspace,
the Regge action factorizes into a sum of independent terms, one
for each of the boundary components, viz.,
\be
I(z^{(1)},...,z^{(N)},S^{(1)},...,S^{(N)}) = \sum_{j=1}^{N}
I(z^{(j)},S^{(j)})\;\;.
\label{r56}
\ee
This result follows from the fact that all $4$-simplices in
the cone are of the form $[c,\sigma_{3}^{(j)}]$, where
$\sigma_{3}^{(j)}$ is a $3$-simplex in the boundary component $M_{3}^{(j)}$.
This ensures factorization of the cosmological term in the action.
Similarly, all internal $2$-simplices are given by $\sigma_{2}(int)=[c,
\sigma_{1}^{(j)}]$, where $\sigma_{1}^{(j)}$ is a $1$-simplex in
$M_{3}^{(j)}$.
The only communication between the different boundary components is
via the cone vertex $c$; for example, there are  no $1$-simplices
of the type $[x,y]$, with the vertices $x$ and $y$ belonging
to different $M_{3}^{(j)}$. As a result, the Einstein term factorizes.
Finally, all boundary $2$-simplices are of the type $\sigma_{2}^{(j)}$,
and the result is secured.

Consequently, the topology changing amplitude factorizes into a product
\be
\Psi_{0}(S^{(1)},...,S^{(N)}) = \prod_{j=1}^{N} \Psi_{0}(S^{(j)})\;\;,
\label{r57}
\ee
where each of the terms in the factorized product is precisely
the wave function of a single component universe with topology
$M_{3}^{(j)}$, which we have already studied.
Thus, we see that within this choice of simplicial minisuperspace,
the basic building blocks of all topology changing amplitudes are
the single component wave functions.
Of course, such a factorization property is a direct consequence of
the restrictive nature of the minisuperspace, and the simplicity of the
cone type cobordism. Nevertheless, within the Regge calculus approach,
such structures are indeed allowed.
It is straightforward now to compute these topology changing amplitudes
which involve a number of lens spaces of varying topology, by using
the explicit form of the wave functions of the previous section.

\section{Behaviour Under Boundary Subdivision}

The final topic of our investigation is to determine how the above
results behave when the boundary universe undergoes a simplicial
subdivision. While the triangulations presented in section $3$ do
indeed capture the topology of the lens spaces, they are by no means
the only available triangulation. Indeed, one can subject these
complexes to various subdivision moves, which yield combinatorially
equivalent triangulations.

For our purposes here, we shall appeal to a set of moves due
to Pachner \cite{Pach}; these are known as $(k,l)$ subdivision moves,
since they replace a set of $k$ simplices by a set of $l$ simplices.
Of particular relevance  to us is the fact that the triangulations
of section of $3$ are simplicial manifolds.
According to the result
of \cite{Pach}, the $(k,l)$ moves are equivalent to
the so-called Alexander moves \cite{Alex} for closed simplicial manifolds.
Therefore,
all combinatorially equivalent triangulations of the lens spaces
can be obtained via the $(k,l)$ moves.

In the three dimensional case of interest here, there are
four $(k,l)$ moves, with $k=1,\cdots,4$, and $k+l=5$.
The $(1,4)$ move is described by adding a new vertex
$x$ to the centre of the
$3$-simplex $[0,1,2,3]$, and linking it to the other $4$ vertices.
The original $3$-simplex is then replaced by  four
$3$-simplices,
\be
[0,1,2,3] \rightarrow
[x,1,2,3] - [x,0,2,3] + [x,0,1,3] \non\\
- [x,0,1,2] \;\;.
\label{r58}
\ee

The $(2,3)$ move involves replacing  two $3$-simplices which share a
common $2$-simplex $[0,1,2]$  by three $3$-simplices sharing a common
$1$-simplex $[x,y]$:
\bea
[x,0,1,2] &-& [y,0,1,2] \rightarrow
[x,y,1,2] - [x,y,0,2] \non\\
&+&[x,y,0,1] \;\;.
\label{r59}
\eea
The $(3,2)$ and $(4,1)$ moves are inverse to the above.

If we denote by $\Delta N_{i}$ the increase in the
number of $i$-simplices due to a $(k,l)$ move, then it is
straightforward to check that under the $(1,4)$ move we have:
\bea
\Delta N_{0} &=& 1\;\;,\non\\
\Delta N_{1} &=& 4\;\;,\non\\
\Delta N_{2} &=& 6\;\;,\non\\
\Delta N_{3} &=& 3\;\;.
\label{r60}
\eea
The changes under the $(2,3)$ move are given by
\bea
\Delta N_{0} &=& 0\;\;,\non\\
\Delta N_{1} &=& 1\;\;,\non\\
\Delta N_{2} &=& 2\;\;,\non\\
\Delta N_{3} &=& 1\;\;.
\label{r61}
\eea

We recall that the Regge action in our simplicial minisuperspace
was fixed in terms of the coefficients $a_{i}$, which in turn were
related to the $N_{i}(M_{3})$.
When we perform a subdivision move of type $(k,l)$ on our
bounding complex, the resulting four dimensional complex is still
a cone over this subdivided boundary. However, the subdivision
creates a number of new internal simplices, and also of course
additional boundary simplices. If we now decide to maintain
the nature of our simplicial minisuperspace, described entirely
in terms of a single internal edge length, and a single
boundary edge length, then we must endow the newly generated edges
with those values.

With this declaration in place, the effects of the subdivision moves
are easily established, the net result being an alteration in the values of
the $a_{i}$ coefficients. The changes can readily
be extracted by combining equations (\ref{r60}) and (\ref{r61})
with equation (\ref{r33}). In this way, we immediately obtain the Regge
action for the cone with subdivided boundary, and most importantly
the analytic structure of the action remains intact,
so that the previous analysis carries through.

One interesting feature of these subdivisions is their influence
on the critical value of the boundary edge length $S_{crit}$.
If we perform $n$ subdivision moves of type $(1,4)$, we find
\be
S_{crit} \rightarrow  4\sqrt{2} \frac{(N_{1}(M_{3})+ 4 n)}
{(N_{3}(M_{3}) + 3n)}
[2 \pi - \frac{6(N_{3}(M_{3}) + 3n)}{(N_{1}(M_{3}) + 4n)}
\cos^{-1}(\frac{1}{3})]\;\;.
\label{r62}
\ee
In the limit of large $n$, the critical value tends to $S_{crit} = 5.6106$.
For the case of $S^{3}$ boundary, for example, this means one can extend
the Lorentzian regime of extrema to smaller values of
the boundary edge length, by performing such subdivisions.

Under a set of $m$ moves of type $(2,3)$, one finds that
\be
S_{crit} \rightarrow  4\sqrt{2} \frac{(N_{1}(M_{3})+ n)}
{(N_{3}(M_{3}) + n)}
[2 \pi - \frac{6(N_{3}(M_{3}) + n)}{(N_{1}(M_{3}) + n)}
\cos^{-1}(\frac{1}{3})]\;\;.
\label{r63}
\ee
The limiting value for large $m$ in this case  is $S_{crit} = -6.23708$.
Therefore, by performing such moves we can ensure that the region
of Lorentzian  extrema covers the full range of physically allowed
values of $S$.

We also note that the factorization property is preserved under
subdivision of the boundary once the nature of the minisuperspace
is kept intact.

\section{Conclusions}

We have studied the ground state wave function  for a universe which
is topologically a lens space. The crucial element in the construction
was to assume that the four dimensional spacetime was modelled by
the cone over the bounding lens space.
By restricting attention to a simplicial minisuperspace, the Regge action
simplified  to the extent that one could investigate its relevant
properties explicitly. Indeed, it should be observed that the coefficients
$a_{i}$ appearing in the action are determined solely in terms of
the the number of $i$-simplices of the boundary universe $N_{i}(M_{3})$.
This followed as a direct consequence of the choice of cone structure for
the four dimensional simplicial spacetime, and was not reliant
on the boundary universe being of lens space topology.
In particular then, the
analysis presented here can be applied to spatial universes with arbitrary
topology. Given a simplicial complex which models a universe with
some general topology, the Regge action in the minisuperspace
of interest here is completely fixed, and is of the form
given in equations (\ref{r29})-(\ref{r32}), with the values of the
$a_{i}$ coefficients given by equation (\ref{r33}).

Of course, one can consider more sophisticated cobordism structures,
and in particular it would be interesting to perform an analysis
when the four dimensional spacetime is itself a simplicial manifold.

Indeed, since the triangulations of \cite{Brehm} are themselves
simplicial manifolds, they provide an ideal opportunity to study
non-trivial cobordism effects in three dimensions. In particular,
for example, if one removes the link of any vertex from the triangulation
of $L(p,1)$, the resulting structure is a simplicial manifold
with $S^{2}$ boundary. In this way, one could study the wave function
for a universe with $S^{2}$ topology, cobordant to a variety of
lens spaces, and compare notes with the wave function obtained
when one takes the trivial cobordism, i.e., the $3$-disc.

\pagebreak
\begin{center}
{\bf Figure Captions}
\end{center}

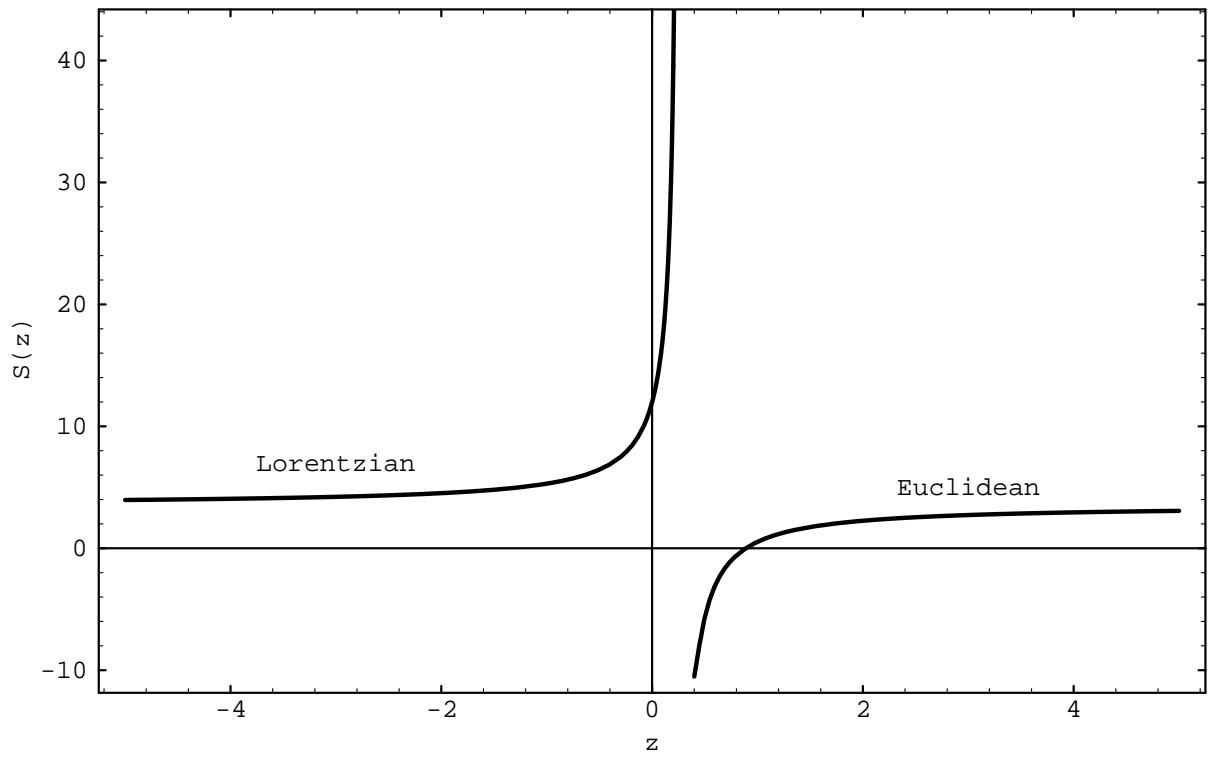
\noindent {\bf Figure 1}: The classical extrema of the Regge action for the
cone over $RP^{3}$.

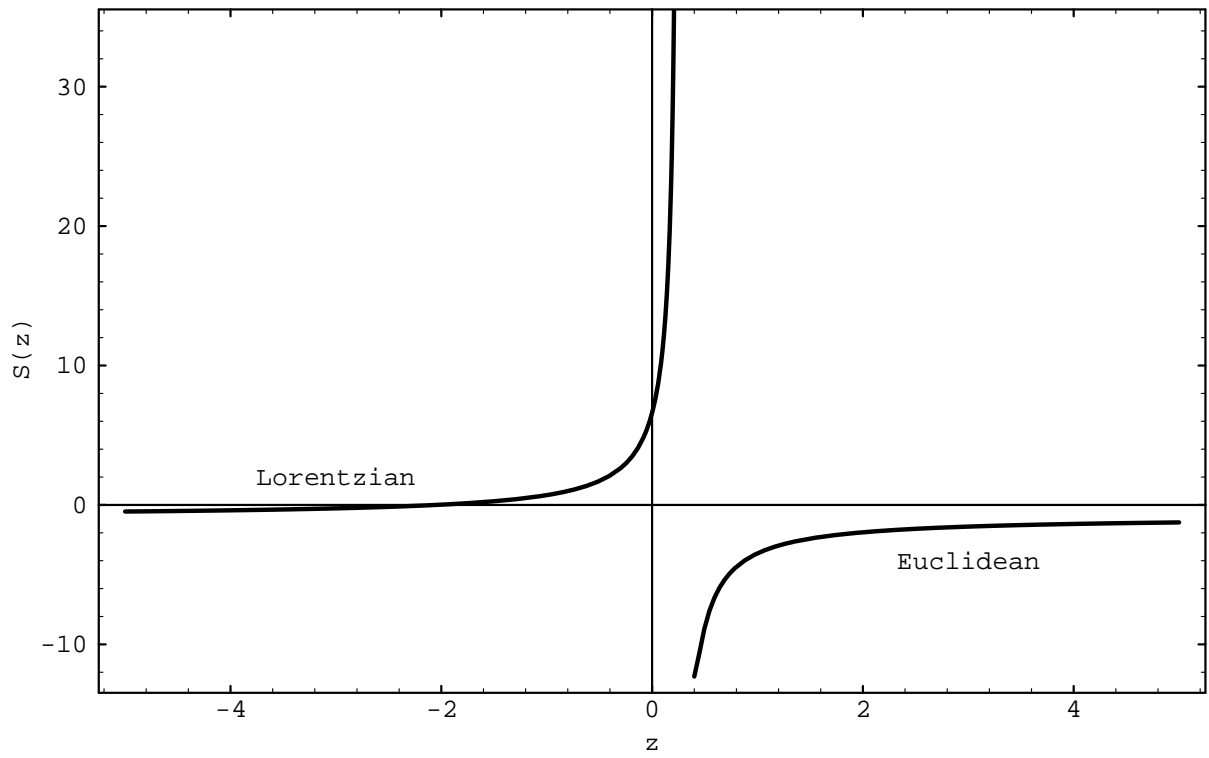
\noindent {\bf Figure 2}: The classical extrema of the Regge action for
the cone over $L(5,1)$. In this case, solutions with Lorentzian signature
are present for all physical values of the scaled
boundary edge length squared $S$. There are no physical Euclidean
signature solutions.

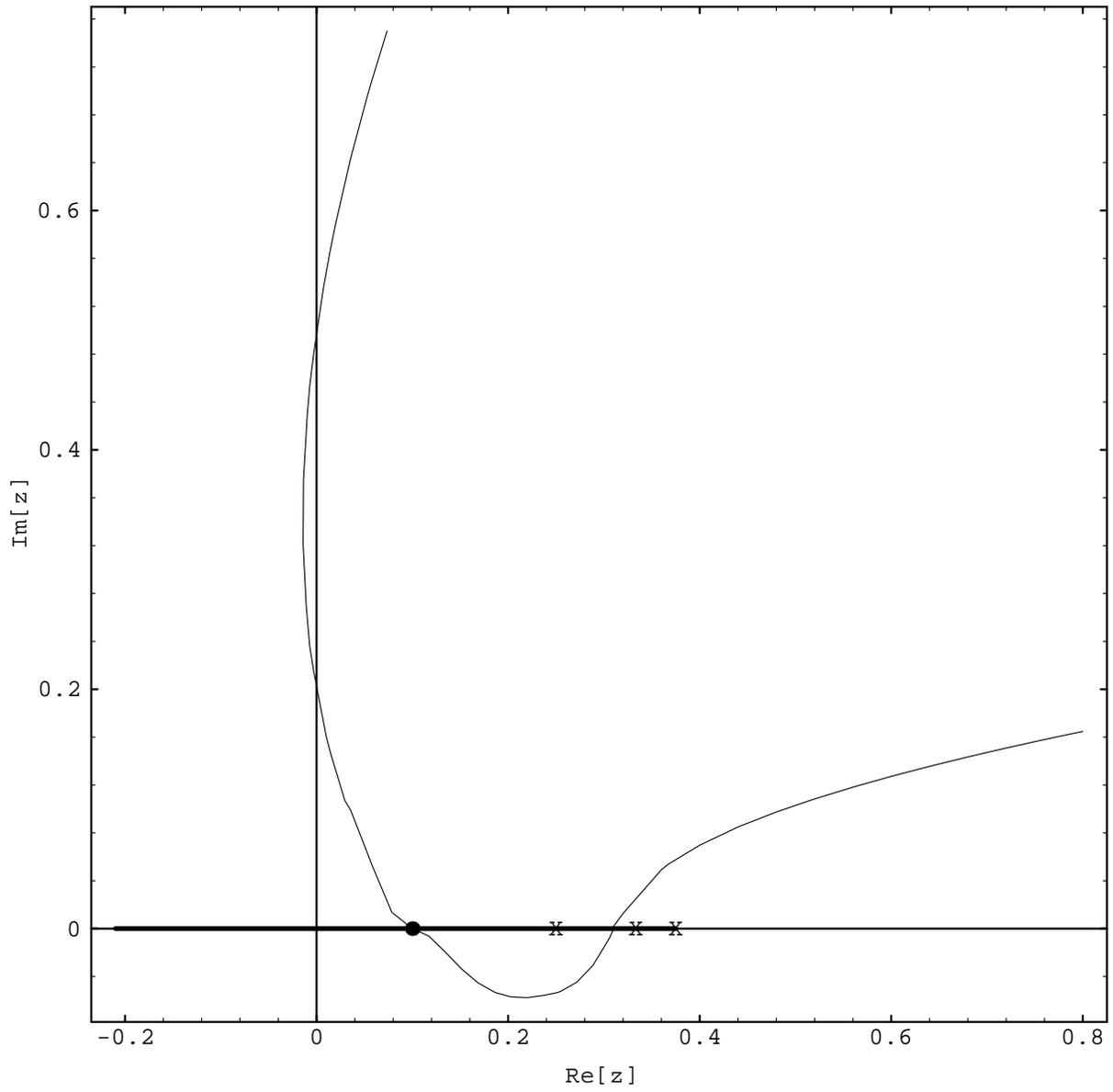
\noindent {\bf Figure 3}: A section of the steepest descent contour of
integration
for the cone over $RP^{3}$. The branch points at $z = \frac{1}{4},\frac{1}{3},
\frac{3}{8}$ are indicated by the crosses, while the Lorentzian extremum at
$Re[z] = 0.1$, corresponding to $S = 17.1246$, is marked by a bold dot.
The value of the action on the first sheet at the extremum is
$I = 1469.32\;i$ and the branch cut is highlighted by the solid
line extending from $z=\frac{3}{8}$ to $-\infty$.
The contour proceeds upwards from the extremum along the first sheet of
the Regge action. Below the branch cut, the contour lies on the second sheet.
Finally, it traverses the cut between the branch points $z=\frac{1}{4}$
and $z= \frac{1}{3}$, reaching a third sheet.

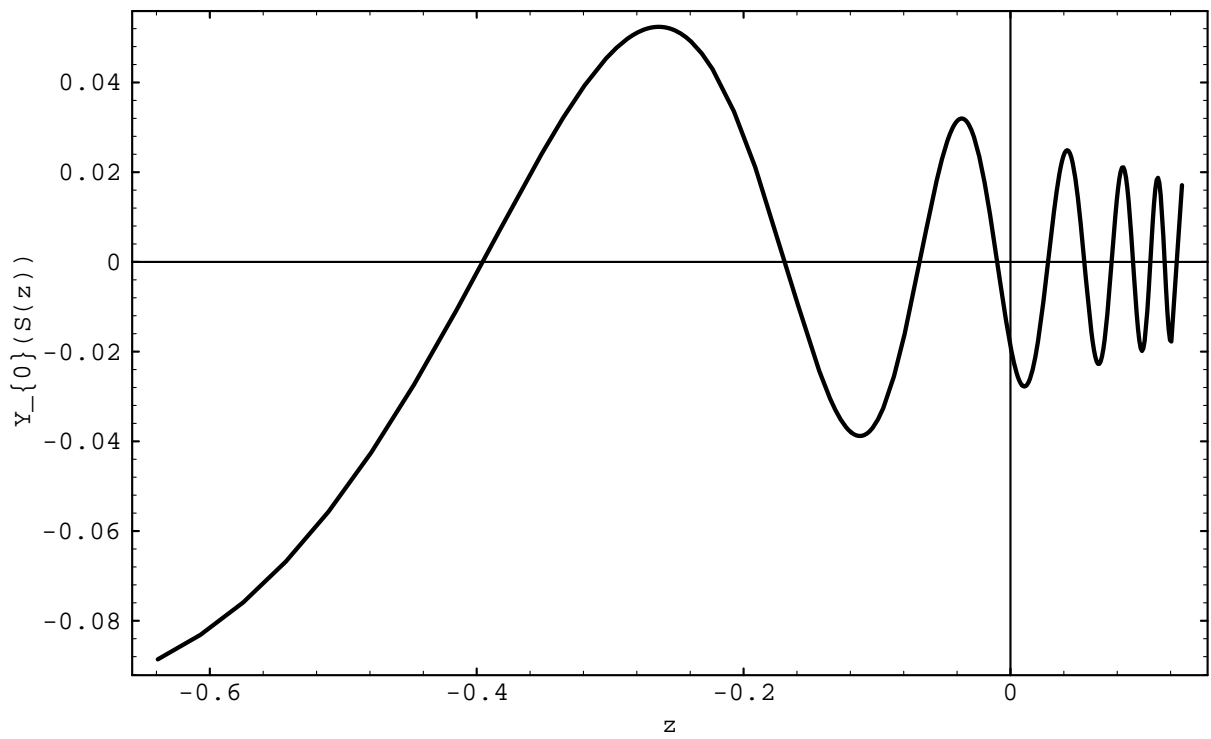
\noindent {\bf Figure 4}: A section of the semi-classical Hartle-Hawking wave
function for a universe with $RP^{3}$ topology.  The range of $S$ values
included is $6 < S < 20$, and $H^{2} = 50$.

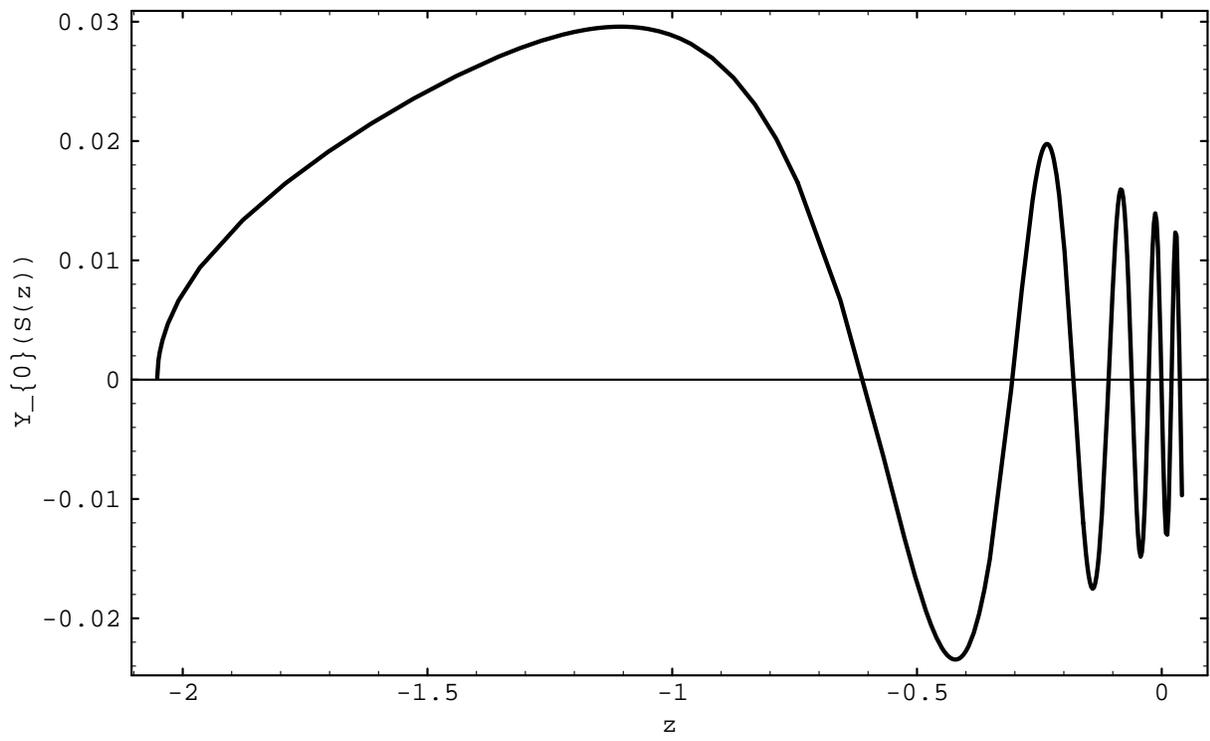
\noindent {\bf Figure 5}: The semi-classical Hartle-Hawking wave function
for a universe with $L(5,1)$ topology, in the range $0 < S < 8$, with
$H^{2} = 50$.

\end{document}